\documentclass[prl,twocolumn,superscriptaddress,showpacs]{revtex4}

\usepackage{amsmath,amsfonts,amssymb,amstext,amsthm}
\usepackage{graphicx}
\usepackage[colorlinks,citecolor=blue]{hyperref}

\def\>{\rangle}
\def\<{\langle}
\def\({\left(}
\def\){\right)}

\newcommand{\ket}[1]{|#1\>}

\newcommand{\braket}[2]{\<#1|#2\>}

\newcommand{\proj}[1]{|#1\>\!\<#1|}

\newcommand{\eref}[1]{Eq.~(\ref{#1})}

\begin{document}

\title{Adiabatic Gate Teleportation}

\author{Dave Bacon}
\affiliation{Department of Computer Science \& Engineering, University of Washington, Seattle, WA 98195}
\affiliation{Department of Physics, University of Washington, Seattle, WA 98195}

\author{Steven T. Flammia}
\affiliation{Perimeter Institute for Theoretical Physics, Waterloo, Ontario, N2L 2Y5 Canada} 

\date{\today}

\begin{abstract}
The difficulty in producing precisely timed and controlled quantum gates is a significant source of error in many physical implementations of quantum computers.  Here we introduce a simple universal primitive, adiabatic gate teleportation, which is robust to timing errors and many control errors and maintains a constant energy gap throughout the computation above a degenerate ground state space.  Notably this construction allows for geometric robustness based upon the control of two  independent qubit interactions.  Further, our piecewise adiabatic evolution easily relates to the quantum circuit model, enabling the use of standard methods from fault-tolerance theory for establishing thresholds.  
\end{abstract}


\maketitle


Building a quantum computer is a daunting task, so much so that it is not even clear which of a plethora of architectures is the most physically viable.  In addition to the standard pulsed implementation of the circuit model of quantum computation (QC), other possible architectures include measurement-based QC~\cite{Raussendorf:01a}, universal adiabatic QC~\cite{Aharonov:04a}, and holonomic QC~\cite{Zanardi:99d}.  Of these, adiabatic QC has recently drawn considerable attention, in part because of its deep connection to computational complexity problems~\cite{Aharonov:04a}, but also due to the advantages this model possesses with respect to decoherence and control~\cite{Childs:01a}.  Similarly holonomic QC has attracted interest because of the geometric robustness of control in this scheme.  Motivated by some of the benefits of adiabatic and holonomic QC, we introduce a new model of QC which is a hybrid between the adiabatic, circuit, and holonomic models.  This model uses nothing but adiabatic quantum evolution, but instead of using a single interpolation between an initial and final Hamiltonian, we use piecewise adiabatic evolutions whose individual parts implement a step in a quantum circuit.  We achieve this by introducing a new primitive: adiabatic gate teleportation (AGT).  

Our route to AGT proceeds by merging two quantum computing protocols: teleportation and adiabatic QC.  Quantum teleportation is the process of transferring the state of a qubit between two distant parties via the use of an initial shared entangled state and two bits of classical communication~\cite{Bennett:93a}.  Notably, while teleportation consumes a Bell pair $|\Phi\rangle=\frac{1}{\sqrt{2}}(|00\rangle+|11\rangle)$ shared between the parties, it can end with a Bell pair localized to the sender.  In adiabatic QC~\cite{Farhi:00a} one adiabatically turns off one Hamiltonian while turning on another Hamiltonian, dragging the system from the ground state of the initial Hamiltonian to that of the final Hamiltonian.  The initial Hamiltonian is chosen such that preparing the system in its ground state can be done efficiently, and the final Hamiltonian is chosen so that its ground state is the solution to a computational problem.  Motivated by teleportation and adiabatic quantum algorithms, we will attempt to adiabatically mimic teleportation.  This will lead us to an adiabatic  protocol for swapping with a simple control scheme that we call adiabatic teleportation.  The main theme of this paper is to use variants on this adiabatic teleportation scheme and the analogy with gate teleportation~\cite{Gottesman:99a} to build a universal quantum computer from piecewise adiabatic evolutions.  Constant-gap piecewise adiabatic evolution \footnote{A constant gap simply means that the minimum gap has no dependence on the total number of qubits.} has previously been considered in the context of state preparation \cite{Schaller:08} and in the context of producing geometric quantum gates from noncyclic adiabatic evolution~\cite{Kult:06a}.  Our model is distinguished from these results by achieving universality and geometric robustness with separately controlled interactions, and by its explicit connection to gate teleportation.  



{\em Adiabatic Teleportation} --- Our setup uses three qubits.  The first qubit is the qubit whose state we wish to transport (swap) to the third qubit.  The second qubit is merely a mediator, which (we will see) is necessary.  At the beginning of the computation we construct a system whose ground state has a single Bell pair $|\Phi\rangle$ on the second and third qubit.  We then adiabatically drag the system to a new Hamiltonian whose ground state has a Bell pair on the first and second qubit (again $|\Phi\rangle$.)  Throughout the evolution the lowest energy level, which is two-fold degnerate, remains degenerate.  If we encode a single qubit of information into this degeneracy, then after this adiabatic evolution the information in this first qubit will now reside in the third qubit.

We choose the initial Hamiltonian for our three qubits to be $H_i=-\omega (X_2X_3+Z_2Z_3)$ and the final Hamiltonian to be $H_f=-\omega (X_1X_2+Z_1Z_2)$ where $X$ and $Z$ are single qubit Pauli matrices, $P_j$ represents the operator $P$ acting on the $j$th qubit, and the identity acting on all other qubits and $\omega$ sets the energy scale.  The ground state of $H_i$ is two-fold degenerate: we can choose a basis for this space as  $|0\rangle \otimes |\Phi\rangle$ and $|1\rangle \otimes |\Phi\rangle$.  Similarly, the ground state of $H_f$ is spanned by $|\Phi\rangle \otimes |0\rangle$ and $|\Phi\rangle \otimes |1\rangle$.  In other words, initially we can store a qubit of information in the first qubit and in the final system we can store it in the third qubit and both configurations are ground states of their respective Hamiltonians.

Now suppose we adiabatically drag the system between $H_i$ and $H_f$.  For example, we may linearly turn off $H_i$ and turn on $H_f$ so that $H(s)=(1-s)  H_i +  s H_f$ from time $s=0$ to $s=1$ and $s=t/T$ is a dimensionless scaled time with scale $T$.  (Other interpolation schemes are certainly possible, and indeed this is one of the benefits of using an adiabatic evolution.)  The above evolution moves the information stored in the first qubit to the third qubit, as we now show.  Let's first define logical qubit operators 
\begin{eqnarray}
	\bar{X}_1&=&XXX,\quad \bar{X}_2=IXX,\quad 
	\bar{X}_3=XXI, \nonumber \\
	\bar{Z}_1&=&ZZZ,\quad ~\bar{Z}_2=ZZI,\quad  ~\bar{Z}_3=IZZ \, .
	\label{eq:encoded}
\end{eqnarray}
Initially we are in the $+1$ eigenstate of $\bar{X}_2$ and $\bar{Z}_3$.  Writing $H(s)$ in this basis we find 
\begin{align}
	H(s)=-\omega (1-s) \big(\bar{X}_2+\bar{Z}_3\big) - \omega s \big( \bar{X}_3+\bar{Z}_2 \big) \, .
\end{align}
Since this Hamiltonian does not include the first logical qubit, it is untouched by the evolution.  This Hamiltonian is nothing more than the time dependent sweeping of $\bar{X}_2$ to $\bar{Z}_2$ and $\bar{Z}_3$ to $\bar{X}_3$.  Evidently this means that if we perform the above evolution slow enough, then, since we start in the $+1$ eigenstates of $\bar{X}_2$ and $\bar{Z}_3$, at the end of the evolution we will be in the $+1$ eigenstates of $\bar{Z}_2$ and $\bar{X}_3$.  A minimum energy gap of $\sqrt{2}\omega$ occurs when $s=1/2$.

Can we figure out what  happens to the first qubit under the above evolution?  We can express the first qubit Pauli operators in terms of the above logical qubits: $ZII=\bar{Z}_1 \bar{Z}_3$ and $XII=\bar{X}_1 \bar{X}_2$.  Since we start off in the $+1$ eigenspace of $\bar{Z}_3$ and $\bar{X}_2$, we see that the logical information is really encoded into the first logical qubit.  As we have argued above, this qubit is untouched by the evolution.  Thus when $s=1$ we must have the same logical information in the first qubit, but now be in the $+1$ eigenvalue subspace of $\bar{Z}_2$ and $\bar{X}_3$.  Now notice that $IIZ=\bar{Z}_1 \bar{Z}_2$ and $IIX=\bar{X}_1 \bar{X}_3$.  Thus we see that actually the information from the first qubit has been dragged to the information on the last qubit.  

Because the gap of the above adiabatic quantum evolution is constant, if we evolve the system sufficiently slowly and in a smooth enough manner, then the adiabatic theorem guarantees that we can achieve the above process with a high fidelity.  There are numerous adiabatic theorems that can be proven~(see for example \cite{Jansen:07a}) which provide guarantees that by making $T$ sufficiently large we can increase the probability that the adiabatic evolution will act successfully (meaning the probability that the system is excited out of the desired subspace is smaller than some constant).  Choosing $T \gg O\left({1 \over \omega}\right)$ is sufficient to guarantee a constant error probability below the threshold for fault-tolerant QC~\cite{Schaller:06a}.


{\em Three Qubits are Necessary} --- We have shown that it is possible to swap quantum information between two qubits via a simple adiabatic interpolation between two fixed Hamiltonians on three qubits.  Is it possible to achieve a similar result without the ancilla qubit?  If we wish to simply interpolate between two two-qubit Hamiltonians, then no.  This does not imply that it is impossible to adiabatically swap two qubits, only that a construction which behaves like the adiabatic quantum algorithm is not possible.  We will also see how this null result implies significant benefits over other adiabatic schemes such as holonomic QC.  

Suppose we have two qubits which we wish to swap by adiabatically ramping between an initial Hamiltonian $H_a$ and a final Hamiltonian $H_b$.  The initial and final Hamiltonians are required to be degenerate such that we can store a single qubit of information in these systems.  Further the initial (final) Hamiltonian must allow for this degeneracy to reside only in the first (second) qubit.  Without loss of generality, we can pick a basis for the first and second qubit so that $H_a$ and $H_b$ are
\begin{eqnarray}
H_a&=&\delta_1 (\proj{01} + \proj{11}) + \delta_2 \proj{00} + \delta_3 \proj{10} ,\nonumber \\
H_b&=&\gamma_1 (\proj{10} + \proj{11}) + \gamma_2 \proj{00} + \gamma_3 \proj{01} ,\nonumber 
\end{eqnarray}
respectively.  Now assume that we turn off $H_a$ while turning on $H_b$.  This leads to the Hamiltonian $H(s)=f(s) H_a + g(s) H_b$, where $f(s)$ ($g(s)$) is a slowly decreasing (increasing) function with $f(0)=1$ and $f(1)=0$ ($g(0)=0$ and $g(1)=1$).  Notice, however, that $H(s)$ is always diagonal in the basis we picked, and therefore the system cannot transform amplitude between these states as required for a swap.  It is crucial here that we assume a simple ramping on and off of the Hamiltonians.  More complicated control schemes lead to holonomic QC which differs significantly from our approach.


{\em Adiabatic Gate Teleportation} --- We have shown how to swap a qubit from the first qubit to the third qubit using adiabatic evolution and now we will show how this can be used to achieve universal QC.  First we will show how in the process of swapping we can also apply a single qubit gate by a simple modification of our initial Hamiltonian.    We label this protocol adiabatic gate teleportation (AGT) in analogy with how gates can be teleported in the quantum circuit model~\cite{Gottesman:99a}.

Suppose, in analogy with the teleportation of quantum gates, that we apply a unitary rotation on the third qubit on the initial Hamiltonian $H_i$: i.e. consider the initial Hamiltonian $H_i^\prime=U_3H_iU_3^\dagger$.  Such an operation does not change the final Hamiltonian, but does change the initial Hamiltonian.  We can then carry the above analysis forward as before, but now in this changed basis.  At the end of the evolution we end up with the logical qubit dragged to the third physical qubit in a rotated basis.  The gap remains $\sqrt 2 \omega$ since the spectrum is unchanged by a unitary conjugation.  Thus it is possible, using this construction, to perform any single-qubit unitary during the adiabatic teleportation.  Notice that the rotated $H_i$ will still consist of two-qubit interactions.  For example, if we wish to perform a Hadmard gate, we can use the same final Hamiltonian, $H_f=-\omega(X_1X_2+Z_1Z_2)$, but chang the initial Hamiltonian to $H_i^\prime=-\omega(X_2 Z_3+Z_2 X_3)$.

It is possible to make different assumptions about how the new, rotated $H_i^\prime$ Hamiltonian arises physically.  We can just assume, for example, that a set of $H_i$ are available in order to perform the desired quantum gates.  A different assumption is that we start with only Hamiltonians of the form $-\omega(X_a X_b + Z_a Z_b)$ between qubits $a$ and $b$, but allows for one to adiabatically drag this Hamiltonian to other ``gate teleporting'' Hamiltonians.  In this model we must ensure that the total system remains in the ground space for the entire evolution, so we must also adiabatically transition from our canonical initial Hamiltonian $H_i=-\omega (X_aX_b+Z_aZ_b)$ to a new Hamiltonian $H_i^\prime=U_b H_i U_b^\dagger$ which leaves the $a$ qubit untouched but prepares $U_b$ on the $b$ qubit for AGT.  We call this adiabatic gate preparation (AGP).  In general, such an evolution isn't directly possible for an arbitrary choice of $U$.  
(For example, consider $U_b = X_b$.)  We can circumvent this by using a universal gate set for a single qubit where every member of the gate set yields an $H(t)$ with a gap.  For instance, we can choose the unitaries
\begin{align}
	A = \frac{1}{2}
	\begin{pmatrix}
		1+i \sqrt{2} & 1 \\
		1 & -1+i \sqrt{2}
	\end{pmatrix} , \,
	B = 
	\begin{pmatrix}
		1 & 0 \\
		0 & e^{i \pi/4}
	\end{pmatrix} ,
\end{align}
which have the requisite properties.  The $A$ matrix is, up to a phase, a square root of the Hadamard matrix, i.e. $A^2 = i (X+Z)/\sqrt{2}$, while $B$ satisfies $B^4 = Z$.  The minimum AGP gaps are $\sqrt{2} \omega$ and $\sqrt{2+\sqrt{2}} \omega$, respectively, at $s=1/2$.  Together, $A$ and $B$ generate $SU(2)$ and hence are universal for single-qubit operations (see pg. 196 of \cite{Nielsen:00a}).


Next consider how to achieve two-qubit gates during the swapping of two qubits.  To do this we follow as above, but instead of applying a single-qubit gate, we apply a two-qubit gate on the final two output qubits.  For example, suppose that we wish to apply a controlled-phase between two logical qubits.  Then we start with  
\begin{align}
	H_i &= -\omega C_Z\big(X_2 X_3+ Z_2 Z_3+ X_5 X_6+ Z_5 Z_6\big)C_Z^\dagger  \nonumber \\
	&= -\omega\big(X_2 X_3 Z_6+Z_2 Z_3+  Z_5 Z_6 + Z_3 X_5 X_6\big), \label{eq:twoqubit}
\end{align}
and end with the Hamiltonian $H_f=-\omega(X_1 X_2 +Z_1 Z_2+X_4 X_5+Z_4 Z_5)$ where  $C_Z$ is the controlled-phase between the the third and sixth physical qubit.  Notice that the gap in this system is again the same constant $\sqrt{2} \omega$, but now we require three-qubit interactions.

We can bypass the inconvenient three-qubit interactions by using perturbation theory gadgets \cite{gadgets, Bartlett:06a}, i.e. two-body Hamiltonians whose low energy dynamics mimic three-qubit interactions.  The price is a reduction in the energy gap by a constant.  In the appendix at the end of this paper we provide a detailed analysis of one such construction.  The crux of this analysis shows that we can use two ancilla qubits and interactions of strength $\omega$ and $\lambda$ to produce an adiabatic evolution with energy gap $O\Big( \frac{\lambda^2}{\omega} \Big)$ with a gate fidelity of $1-\frac{\lambda^2}{2\omega^2}+O \Big( \frac{\lambda^4}{\omega^4} \Big)$.

\begin{figure}[h]
\begin{center}
\includegraphics[scale=0.3]{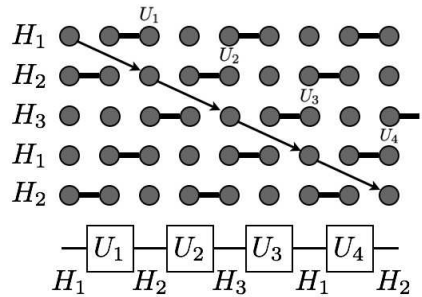}
\caption{By adiabatically dragging, cyclically, between only three Hamiltonians, $H_1$, $H_2$, and $H_3$ we can perform universal quantum computation.  Here we diagram how this works for a single-qubit computation (circuit below, the Hamiltonians at different times diagramed from top to bottom.)  Each circle represents a qubit, and a bar represents a two-qubit Hamiltonian in our scheme, rotated by a labeled unitary $U_i$.  Notice how in each step to the next Hamiltonian, the qubit is swapped over two qubits (the arrows) and a gate is applied to this qubit.  Thus the gates to be applied are encoded spatially across the the three Hamiltonians.  The $i$th gate thus depends on the the Hamiltonian $H_{(i-1)~{\rm mod}~3+1}$ with the gate being applied changing the interaction between qubits $2i$ and $2i+1$ in this Hamiltonian.   Generalizing to more than one qubit this proves that universal holonomic quantum computation can be done by interpolation between only three Hamiltonians. \vspace{-20pt}} 
\label{fig} 
\end{center}
\end{figure}

Putting this all together we have shown how to use AGT to perform one- and two-qubit gates by teleporting quantum information adiabatically between qubits.  Given the ability to prepare fiducial initial single-qubit states and the ability to measure the qubits which contain the state of the final system, we then obtain a model equivalent in power to the standard circuit model of QC.


{\em Relationship to Holonomic QC} --- In holonomic quantum computing (HQC) one uses a cyclic adiabatic evolution of a Hamiltonian around a loop in parameter space to produce a quantum gate.  Almost all HQC is cast within the context of cyclic evolutions, with the exception of Kult {\em et al.} \cite{Kult:06a} who pointed out that noncyclic geometric gates are also possible.  AGT is a example of a noncyclic geometric gate: so long as the evolution is adiabatic and we remain within the control manifold defined by the two interactions we are turning on and off, the desired gate is enacted independent of the actual time dependence of the path taken.  Our construction is distinguished in two ways.  First, we achieve robustness by turning on and off interactions between two different subsystems (as opposed to controlling interactions within the same system), and we expect that the separation of control needed to make geometric evolution robust will be much easier to achieve in this setting.  Second, our explicit connection to gate teleportation leads directly to universal QC and enables methods from fault-tolerance theory. 


{\em Possible Architectures} --- There are many different schemes for using the above AGT primitives to build a universal quantum computer.  Using minimal resources, we can build a circuit on $n$ qubits using only $n+6$ qubits ($4$ qubits for the two extra gates and $2$ for the ancillas in the perturbation gadgets) assuming that we can move the qubits involved in the Hamiltonians around at will.  More realistic and interesting architectures disallow such movement, but allow the parallel circuit elements required for fault-tolerant QC.  

One very compelling architecture builds a circuit on $n$ qubits using $3n$ qubits (plus $n$ ancilla gadget qubits) in a quasi-one-dimensional architecture.  The idea here is simply that one can perform alternating steps in a quantum circuit by gate teleportation from the first $n$ qubits to the third $n$ qubits and then back to the first $n$ qubits.   Another possible architecture builds a quantum circuit of length $l$ on $n$ qubits onto teleportation across $n(2l+1)$ qubits by simply imprinting the quantum circuit being implemented spatially (in a manner similar to what occurs in one-way quantum computing~\cite{Raussendorf:01a}.)  Thus we can perform universal QC by interpolating between just three different fixed Hamiltonians (see Fig.~\ref{fig}.)


{\em Fault Tolerance} --- A full analysis of fault-tolerance in the piecewise adiabatic scheme is beyond the scope of this letter, but here we argue that our system should show similar behavior to fault-tolerance in the standard quantum circuit model.  The reason for this is simply that AGT, while using adiabatic evolution, essentially has the behavior of producing a gate on some (teleported) quantum information.  Thus we could use the standard techniques for proving a threshold on this model.  That said, however, in practice this model may perform significantly better than the standard circuit model.  The reason is that the system is always performing adiabatic evolution with a constant energy gap (unlike many other models which yield energy gaps which scale inversely as a polynomial in the number of qubits.)  Thus we obtain two of the benefits of adibatic QC, (1) the system is separated by a constant energy barrier from, and thus at low temperature is robust to, excitation out of the ground state (a form of leakage error) and (2) considerable robustness exists with respect to varying the tunings which change the Hamiltonian adiabatically.  


{\em Comparison to Other Schemes} --- Using piecewise adiabatic quantum gate teleportations to build a quantum computer shares similarities with many other schemes, but differs in many respects as well.  Like universal adiabatic QC, the scheme uses a smooth one way interpolation between an initial and final Hamiltonian, but we use multiple such interpolations.  Like holonomic QC, we rely on degenerate levels of a Hamiltonian, but here our adiabatic evolution is not cyclic.  Along these lines, our scheme is related to a recent method to make holonomic QC fault-tolerant \cite{Oreshkov:09a} by using interpolations between encoded Pauli operators.  In contrast to our proposal, these are done in a cyclic fashion and with three-qubit interactions.  Further we achieve a gate by controlling interactions between separate subsystems, thus insuring that the geometric robustness depends only on the degree to which these independent controls can be manipulated.

Finally the scheme is similar in spirit to recent proposals to use spin chains with adiabatic time-dependent interactions to transmit quantum information~\cite{Eckert:07a}, where interpolation between two spin-$1$ Hamiltonians was used to transmit quantum information down the chain with an energy gap that scaled (at least numerically) as $1/l$ where $l$ is the length of the chain.  By contrast, our scheme maintains a constant energy gap for the entire computation.  While both schemes require similar transmission times, the former~\cite{Eckert:07a} has a small energy gap, which will be a problem when using this scheme at finite temperature.  Furthermore, by explicitly connecting our scheme to gate teleportation, we achieved a universal QC.


{\em Discussion} --- We have shown how to build a universal quantum computer using a series of piecewise adiabatic quantum evolutions related to teleportation.  This opens up a novel architecture for building a quantum computer based entirely on adiabatic quantum evolutions between two-qubit interactions and it considerably simplifies the control requirements for building a quantum computer.  



After completing this paper we became aware of concurrent work done independently by Oreshkov~\cite{Oreshkov:09b} showing a similar result using cyclic two-qubit interpolations.

\acknowledgments

We thank D.~Gottesman for discussions.  DB was supported by NSF  grants 0803478 and 0829937 and DARPA QuEST grant FA-9550-09-1-0044.  STF was supported by the Perimeter Institute for Theoretical Physics.  Research at Perimeter is supported by the Government of Canada through Industry Canada and by the Province of Ontario through the Ministry of Research~\& Innovation.



\appendix


\section{Isotropic Exchange}

Here we show that if one uses an isotropic exchange interaction instead of the anisotropic interactions in the teleportation protocol one can also perform adiabatic teleportation.  In this case the initial Hamiltonian is
\begin{eqnarray}
H_i &=& \omega (X_2 X_3 +Y_2 Y_3+Z_2 Z_3) \nonumber \\
&=&  \omega (\bar{X}_2+\bar{Z}_3-\bar{X}_2 \bar{Z}_3) \nonumber \\
&=& \omega\left[I-(I-\bar{X}_2)(I-\bar{Z}_3) \right]
\end{eqnarray}
and the final Hamiltonian is
\begin{eqnarray}
H_f &=& \omega (X_1 X_2 +Y_1 Y_2+Z_1 Z_2) \nonumber \\
&=&  \omega (\bar{Z}_2+\bar{X}_3+\bar{Z}_2 \bar{X}_3) \nonumber \\
&=& \omega\left[I-(I-\bar{Z}_2)(I-\bar{X}_3)   \right]
\end{eqnarray}
where we have expressed these Hamiltonians in terms of the encoded operations given in Eq. 1 of the main text.  These equations show that now instead of two decoupled encoded qubits, the encoded qubits are coupled.  However notice that the initial ground state is the $-1$ eigenstate of $\bar{X}_2$ and $\bar{Z}_3$ and the final ground state is the $-1$ eigenstate of the $\bar{Z}_2$ and $\bar{X}_3$, just as in anisotropic exchange protocol, but with the signs flipped.  Further there are no level crossing in a linear ramping between these two Hamiltonians, and the gap is a constant $2 \omega$ occurring at the midpoint of this evolution.  Thus the adiabatic teleportation protocol caries through for the isotropic exchange.  Notice, importantly, that the coupling however must be antiferromagnetic.


\section{Three-qubit effective interactions}

Here we provide more details on how to implement the Hamiltonian in Eq. 4 in the main text using the perturbation theory gadgets of Bartlett and Rudolph~\cite{Bartlett:06a}.  In these gadgets, one replaces one of the qubits in a three-qubit interaction by an {\em encoded} qubit across two qubits.  Since we need two three-qubit interactions, this means that we require two extra qubits in this construction.  We label our logical qubits $L$ and $R$ (for left and right), and encode each into four physical qubits labeled 1 -- 4.  

\begin{figure}[t]
\begin{center}
\includegraphics[scale=0.45]{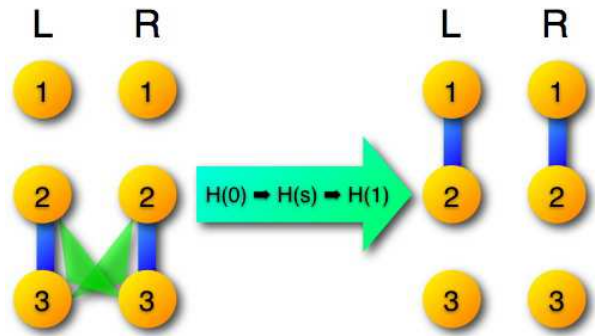}
\caption{Ideal two-qubit adiabatic gate teleportation using three-body Hamiltonian interactions. There are two logical qubits, $L$ and $R$, each encoded in three physical qubits $1,2,3$, where the ancillas facilitate the teleportation as discussed in the main text.  Blue bars represent $ZZ$ couplings, while green triangles represent interactions of the form $X^L X^L Z^R$ (and similarly with $L\leftrightarrow R$) as in \eref{E:ideal}. \vspace{-20pt}} 
\label{F:ideal}
\end{center}
\end{figure}

Let's recall Eq. 4, relabeled here as in Fig.~\ref{F:ideal}.  The ideal initial Hamiltonian is
\begin{align}\label{E:ideal}
	H_{\rm ideal} = -\lambda \big(X_2^L X_3^L Z_3^R + Z_2^L Z_3^L \big) + [L \leftrightarrow R] ,
\end{align}
where $[L \leftrightarrow R]$ just means to add the terms which exchange the qubits $L$ and $R$.  Now let's add the ancilla qubits and move to the encoded subspace.  The encoded subspaces we are working in are the subspaces spanned by $|00\rangle$ and $|11\rangle$ on qubits $3$ and $4$.  We can force the joint state of qubits 3 and 4 to lie in this subspace by adding a strong $Z_3Z_4$ coupling term to the ideal Hamiltonian.  Thus, \eref{E:ideal} can be realized using encoded operators as the following target Hamiltonian
\begin{align}
	H_{\rm target}=& -\lambda \big(X_2^L \overline{X_3^L}\, \overline{Z_3^R} + Z_2^L \overline{Z_3^L}\big) \nonumber \\ 
	&  -\omega\big(Z_3^L Z_4^L\big) + [L \leftrightarrow R]. 
\end{align}
Here the encoded operators (with bars on top) are
\begin{align}
	\overline{X_3} = X_3 X_4 \quad , \quad \overline{Z_3} = Z_3 \mbox{ or } Z_4 \, ,
\end{align}
for both the left and right qubits and we are assuming that the coupling strengths satisfy $\omega \gg \lambda$.  We are free to choose either $Z_3$ or $Z_4$ for the encoded $\overline{Z_3}$ operation because these operators act equivalently up to multiplication by the stabilizer of the encoded subspace.  Writing this out in terms of the Pauli operators on the physical qubits, we find (for one such choice of encoded $Z$)
\begin{align}
	H_{\rm target}=&-\lambda \big(X_2^L X_3^L X_4^L Z_4^R + Z_2^L Z_3^L \big) \nonumber \\ & -\omega\big(Z_3^L Z_4^L\big) + [L \leftrightarrow R] .
\end{align}
Following  Bartlett and Rudolph, we use the following initial Hamiltonian.  It is a two-body gadget Hamiltonian that simulates the low energy behavior of the above target Hamiltonian, and is given by
\begin{align}
	H_i =& -\lambda \big(X_2^L X_3^L + X_4^L Z_4^R + Z_2^L Z_3^L\big) \nonumber \\
&-\omega \big(Z_3^L Z_4^L\big) + [L \leftrightarrow R] .
\end{align}
The $\omega$ term in this Hamiltonian by itself would force the ground state of qubits $3$ and $4$ to be in the subspace spanned by $|00\rangle$ and $|11\rangle$ as discussed above.  The $\lambda$ term is now a two-qubit interaction which simulates the four-body term in the target Hamiltonian.  

\begin{figure}[t]
\begin{center}
\includegraphics[scale=0.45]{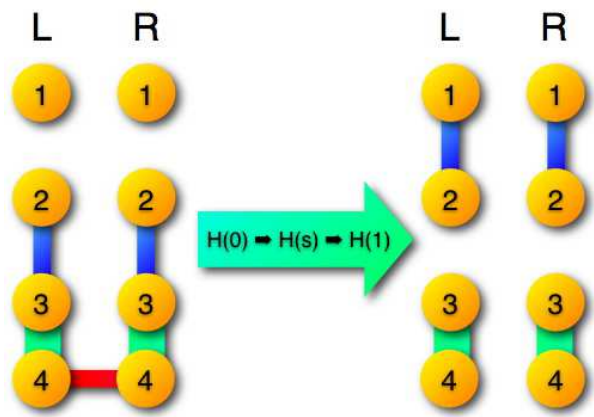}
\caption{Two-body interactions can simulate the ideal three-body process of Fig.~\ref{F:ideal}.  The simulation gadget uses one additional ancilla qubit (labeled $4$) per logical qubit.  Qubits $3$ and $4$ are bound by strong $-\omega ZZ$ couplings for the duration of the evolution, as shown by the broad green bars.  Blue bars represent two-body interactions $- \lambda (XX+ZZ)$, while the red horizontal bar represents a $- \lambda (XZ+ZX)$ coupling. When the coupling strengths are chosen so that $\lambda \ll \omega$, this adiabatic evolution simulates the ideal evolution of \eref{E:ideal} and Fig.~\ref{F:ideal}.  The fidelity of the simulation is $1-O(\lambda^2/\omega^2)$ and the energy gap governing the adiabatic condition is given by $\Delta E  \ge \lambda^2/\omega$. \vspace{-20pt}} 
\label{F:gadget}
\end{center}
\end{figure}

Our desired final Hamiltonian is given by
\begin{eqnarray}
H_f&=& -\lambda \big( X_1^L X_2^L +Z_1^L Z_2^L  \big) \nonumber \\
 &&-\omega \big( Z_3^L Z_4^L \big) + [L \leftrightarrow R] .
\end{eqnarray}
Notice, importantly, that we leave on the interaction which forces qubits $3$ and $4$ into the encoded subspace during the entire evolution.  As usual, our total evolution is given in terms of the scaled time $s=t/T$ by
\begin{align}
	H(s) = (1-s) H_i + s H_f .
\end{align}
This evolution is depicted in Fig. \ref{F:gadget}.

We must show that the above adiabatic evolution has high fidelity with the ideal evolution and that the gap is not too small.  The fidelity is governed by the overlap of the ground state of $H_i$ with the ground state of the ideal (encoded) Hamiltonian $H_{\rm target}$.  

In order to analyze this gadget it is useful to perform a change of basis.  In particular if one undoes the controlled-phase gate between the ancilla qubits $4L$ and $4R$, then above Hamiltonian becomes a sum of terms acting separately on $L$ and $R$.  Since this is a unitary conjugation it doesn't change the gap, and we can also find the ground state in this basis and transform back.  Since the Hamiltonian is now decoupled across $L$ and $R$, we drop these subscripts now and write the transformed initial Hamiltonian as
\begin{align}
	H_i^\prime =& -\lambda (X_2 X_3 + X_4 + Z_2 Z_3)-\omega (Z_3 Z_4) .
\end{align}
Note that the final Hamiltonian is unaffected by this transformation, and so we need merely to drop the $L$ and $R$ superscripts to obtain
\begin{align}
	H_f^\prime =& -\lambda (X_1 X_2 + Z_1 Z_2)-\omega (Z_3 Z_4) .
\end{align}

Let's first find the ground state of the initial Hamiltonian so we can check the fidelity.  We can further simplify things by applying a controlled-not gate from qubit 3 to qubit 2 resulting in
\begin{align}\label{E:twoprimes}
	H_i^{\prime\prime} =& -\lambda Z_2 -\lambda (X_3 + X_4)-\omega (Z_3 Z_4) .
\end{align}
In this basis, qubits 1 and 2 completely decouple, and the fidelity depends only on the overlap of this ground state with the Bell state $\ket{\Phi} = \frac{1}{\sqrt{2}}(\ket{00}+\ket{11})$ on qubits 3 and 4.  We can exactly diagonalize by first transforming to the Bell basis.  Let's define $r=\lambda/\omega$ to be our small expansion parameter.  Then the ground state of \eref{E:twoprimes} on qubits 3 and 4 is given by
\begin{equation}
\ket{g} = \alpha \frac{1}{\sqrt{2}}(|00\rangle+|11\rangle) + \sqrt{1-\alpha^2}\frac{1}{\sqrt{2}} (|01\rangle +|10\rangle)
\end{equation}
where the coefficient $\alpha$ is
\begin{equation}
	\alpha=\(\frac{1}{2}+\frac{1}{2 \sqrt{4 r^2+1}}\)^{1/2}.
\end{equation}
Expanding in powers of $r=\lambda/\omega$, the fidelity is
\begin{align}
	|\braket{\Phi}{g}| = \alpha = 1- {\lambda^2 \over 2\omega^2} + O\bigg({\lambda^4 \over \omega^4}\bigg) ,
\end{align}
which is corrected at second order in $r$.

Now let's compute the gap to see what price we must pay to achieve high fidelity.  In the basis where we have applied a controlled-not from qubit 3 to qubit 2, the final Hamiltonian is
\begin{equation}
H_f^{\prime \prime}=- \lambda (X_1 X_2 + Z_1 Z_2 Z_3) -\omega (Z_3 Z_4)
\end{equation}  
Note that $Z_1 Z_2$ and $X_1X_3 X_4$ commute with both the initial and final Hamiltonian, corresponding to the encoded quantum information.  Suppose we work in a basis where this information is in the $+1$ eigenstate of $Z_1 Z_2$.  Then the final Hamiltonian simplifies to
\begin{equation}
H_f^{\prime \prime}=- \lambda (X_1 X_2 + Z_3) -\omega (Z_3 Z_4).
\end{equation}  
Here we see that qubits $1$ and $2$ are decoupled from those of $3$ and $4$.  If one linearly sweeps between these initial and final Hamiltonians, one will obtain a minimal gap for each of these evolutions.  The smaller of these gaps comes from qubits $3$ and $4$.  Explicitly, the evolution to consider is
\begin{equation}
H(s)=-(1-s)[\lambda (X_3 + X_4)+\omega (Z_3 Z_4)] -s[ \lambda  Z_3 +\omega (Z_3 Z_4)].
\end{equation}
The gap between the lowest two eigenvalues of this evolution is
\begin{eqnarray}
\Delta E(s)&=&\omega \sqrt{1+r^2 (3s^2-4s+2)+\chi} \nonumber \\
&&-\omega \sqrt{1+r^2 (3s^2-4s+2)-\chi}
\end{eqnarray}
where 
\begin{equation}
\chi=2\sqrt{r^2s^2+r^4(1-s)^2(2s^2-2s+1)} .
\end{equation}
Using the fact that $\sqrt{1+x}-\sqrt{1-x} \geq x$ for $0 \leq x \leq 1$ we can bound this as
\begin{equation}
\Delta E(s) \geq  \omega  {\chi \over \sqrt{1+  r^2 (3s^2-4s+2)}}.
\end{equation}
We can upper bound the lower equation by $\sqrt{1+2r^2}$, and we can use $(1-s)^2(2s^2-2s+1) \geq (1-s)^3/3$ for $0 \leq s \leq 1$ to express the gap as
\begin{equation}
\Delta E(s) \geq {2 \omega  \over \sqrt{1+2r^2} } \sqrt{r^2s^2+r^4 (1-s)^3/3}.
\end{equation}
This obtains its max at
\begin{equation}
s\ = \frac{1+r^2-\sqrt{1+2 r^2}}{r^2}.
\end{equation}
For $r<0.5$ this yields a bound on the gap of
\begin{equation}
\Delta E(s) \geq \omega r^2
\end{equation}

Thus we have shown that the initial fidelity with the proper ground state is high $(1-O(\lambda^2/\omega^2))$, and also that the energy scale which sets the adiabatic condition is set by the perturbative energy scale, $O(\lambda^2/\omega)$.  For fault-tolerance we require a fixed accuracy and our results imply that the gadget construction can achieve this, albeit at the cost of the energy gap shrinking and thus a slower adiabatic gate time.

\end{document}